\documentclass[aps,pre,twocolumn,superscriptaddress]{revtex4-1}
\usepackage{graphicx}
\usepackage{amssymb}
\usepackage{amsmath}
\usepackage{kotex}
\usepackage[mathlines]{lineno}
\usepackage{color}

\usepackage{xcolor}

\newcommand{\skb}[1]{\textcolor{brown}{#1}}

\usepackage{multirow}
\usepackage{chngcntr}

\let\oldsubequations\subequations
\let\oldendsubequations\endsubequations

\renewenvironment{subequations}
  {\linenomathNonumbers\oldsubequations}
  {\oldendsubequations\endlinenomath}

\begin{document}
\title{Exact cluster dynamics of indirect reciprocity in complete graphs}
\author{Minwoo Bae}
\affiliation{Department of Physics, Pukyong National University, Busan 48513, Korea}
\author{Takashi Shimada}
\email[]{shimada@sys.t.u-tokyo.ac.jp}
\affiliation{Department of Systems Innovation, Graduate School of Engineering, The University of Tokyo, 7-3-1 Hongo, Bunkyo-ku, Tokyo 113-8656, Japan}
\author{Seung Ki Baek}
\email[]{seungki@pknu.ac.kr}
\affiliation{Department of Scientific Computing, Pukyong National University, Busan 48513, Korea}

\begin{abstract}
Heider's balance theory emphasizes cognitive consistency in assessing others, as is expressed by ``The enemy of my enemy is my friend.''
At the same time, the theory of indirect reciprocity provides us with a
dynamical framework to study how to assess others based on their actions as well
as how to act toward them based on the assessments. Well known are the ``leading
eight'' from L1 to L8, the eight norms for assessment and action to foster cooperation in social dilemmas while resisting the invasion of mutant norms prescribing alternative actions.
In this work, we begin by showing that balance is equivalent to stationarity of dynamics only for L4 and L6 (stern judging) among the leading eight. Stern judging reflects an intuitive idea that good merits reward whereas evil warrants punishment.
By analyzing the dynamics of Stern Judging in complete graphs, we prove that this norm almost always segregates the graph into two mutually hostile groups as the graph size grows.
We then compare L4 with stern judging:
The only difference of L4 is that a good player's cooperative action toward a bad one is regarded as good. This subtle difference transforms large populations governed by L4 to a ``paradise'' where cooperation prevails and positive assessments abound.
Our study thus helps us understand the relationship between individual norms and their emergent consequences at a population level, shedding light on the nuanced interplay between cognitive consistency and segregation dynamics.
\end{abstract}

\maketitle


Heider's balance theory is a minimal model of consistency in human relations~\cite{heider1946attitudes}. Because of its mathematical simplicity, the balance theory has opened up an interdisciplinary field between psychology and graph theory~\cite{harary1953notion,cartwright1956structural}. Between each given pair of individuals $i$ and $j$, the theory assumes a mutual binary relation, which can be represented either by $\sigma_{ij} = \sigma_{ji} = +1$ if positive (e.g., they like each other) or by $-1$ if negative (e.g., they dislike each other).
For any graph with positive and negative edges, we say that it is balanced if
and only if all paths joining the same pair of vertices have the same sign,
where the sign of a path is defined as the product of the values of its
constituent edges. If it is unbalanced, one feels ``tension'' because, e.g., one
has a positive relation with his or her friend's enemy so that two paths
connecting the same pair of individuals have different signs.
Of particular importance is the structure theorem~\cite{harary1953notion,cartwright1956structural}, which states that a graph is balanced if and only if the graph is separated into two clusters in such a way that every edge inside a cluster is positive whereas any edge between the clusters is negative. Let us define the size of a cluster as the number of vertices inside it. In a balanced configuration, the size of one of the clusters can be zero, in which case, all the edges are positive, and such a configuration is called paradise~\cite{antal2005dynamics}.

The balance theory is static, and one could naturally ask which dynamics leads to a balanced configuration. The first attempt was a discrete-time model, in which a randomly chosen edge is flipped to achieve more balance, either locally or globally~\cite{antal2005dynamics}.
One version of this discrete-time model corresponds to an Ising-spin system with the following Hamiltonian: $H = -\sum_{\left< ijk \right>} \sigma_{ij} \sigma_{jk} \sigma_{ki}$,
where the summation runs over all triangles of the underlying graph~\cite{malarz2022mean,woloszyn2022thermal,malarz2023thermal}. By making thermal contact with a heat bath at temperature $T$, this Hamiltonian system undergoes stochastic time evolution, which was expected to guide the system to a balanced configuration as $T \to 0$.
However, it has turned out that such local dynamics also generates infinitely many stable yet unbalanced configurations~\cite{marvel2009energy}. An alternative, continuous-time model with real-valued $\sigma_{ij}$'s has thus been studied to prove that a balanced configuration is almost always reached from a random initial configuration in finite time, if a large number of vertices are connected by a complete graph~\cite{kulakowski2005heider,marvel2011continuous}.

In this work, we study dynamics of indirect reciprocity as another way to approach structural balance. An individual's deed to another can be reciprocated indirectly from a third-party observer, and this mechanism is regarded as a powerful mechanism of establishing cooperation among individuals~\cite{alexander1987biology,nowak2005evolution}.
The mechanism of indirect reciprocity typically involves three persons: An
actor, also called a donor, and a recipient of the donation from the donor, and
an observer watching those two persons and updating his or her own assessment of
the donor. The donor may either cooperate by choosing donation or defect by
giving nothing to the recipient.
The combination of an assessment rule and an action rule defines a
norm.
Note that 
$\sigma_{ij}$ does not generally equal $\sigma_{ji}$.

An early issue of debate in indirect reciprocity was whether a first-order assessment rule, relying only on the donor's action, is sufficient for stabilizing the paradise~\cite{nowak1998evolution,leimar2001evolution,sugden1986economics}, and the answer, at least in theory, is that the observer should use information of the recipient as well, in such a way that refusing to cooperate to a bad recipient does not damage the donor's image to the observer~\cite{ohtsuki2004should,ohtsuki2006leading}.
Among such good norms with higher-order assessment rules, which are now called
``leading
eight'', a particularly well-known example is what we will call
L6~\cite{brandt2007survey,hilbe2018indirect} throughout this work (also known as
``stern judging'')~\cite{kandori1992social,pacheco2006stern,santos2018social}. According to this simple and intuitive norm, a donor must donate only when he or she regards the recipient as good, and whether the donor's cooperation looks good to an observer heavily relies on the recipient's image to the observer.
The resulting dynamics of L6 can be algebraically expressed
as~\cite{oishi2021balanced}
\begin{equation}
    \sigma_{od}'=\sigma_{or}\sigma_{dr},
    \label{eq:L6}
\end{equation}
where the prime means an updated value at the next time step, and the subscripts $d$, $r$, and $o$ mean the donor, recipient, and observer, respectively.
Note that Eq.~\eqref{eq:L6} has combined an assessment rule and an
action rule as was originally proposed in Ref.~\cite{ohtsuki2004should}.
L6 has its own weaknesses: Once it deviates from the paradise, L6 fails to
return in the presence of noisy and private
assessment~\cite{uchida2010effect,hilbe2018indirect,lee2021local,lee2022second,mun2023second},
and L6 would divide a fully connected society into two antagonistic
clusters~\cite{oishi2013group}.
By contrast, if the society uses
L4~\cite{brandt2007survey,hilbe2018indirect,santos2021complexity}, which is identical to L6
except that
a good donor's cooperation with a bad recipient
does not damage the donor's image to observers,
our numerical simulations show that the society reaches the paradise.
This work will provide analytic understanding of those numerically observed
phenomena in L6 and L4.

Let us consider a directed graph with a set of vertices and a set of edges, which are denoted as $V$ and $E$, respectively. Each vertex has an agent, who has an opinion about everyone else, including him or herself.
The cardinality of $V$ is the total number of vertices, or the population size, and we denote it as $N$.
Opinions take discrete values, so $\sigma_{ij}=+1$ if $i$ regards $j$ as good, and $-1$ otherwise.
Everyone shares an assessment rule and an action rule in common.
Then, the opinions are updated in the following way:
\begin{enumerate}
    \item A donor $d$ is randomly chosen from $V=\{v_1, \ldots, v_N\}$, and a recipient $r$ is chosen among $d$'s neighbors including $d$ itself.
    \item The donor $d$ chooses whether to cooperate toward the recipient $r$ according to the action rule.
    \item All neighbors of both $d$ and $r$ observe $d$'s action toward $r$. The
    set of observers includes $d$ and $r$ as well. Every observer $o$'s
    opinion about $d$, denoted as $\sigma_{od}$, is updated to $\sigma_{od}'$ as
    prescribed by the given assessment rule with probability $1-\epsilon$, and
    to $-\sigma_{od}'$ with probability $\epsilon\ll 1$ by
    assessment error. Error in implementing an intended
    action will be ignored because it makes no qualitative differences~\cite{suppl}.
\end{enumerate}

We begin by checking whether the balance condition is equivalent to stationarity for each of the leading eight. This analysis is necessary because it relates the balance condition, a static property, to stationarity, which is a dynamical consequence.
Regarding L6, we may rewrite its rule [Eq.~\eqref{eq:L6}] as
\begin{equation}
\sigma_{od}'=\sigma_{or}\sigma_{dr}=\sigma_{od}\sigma_{or}\sigma_{dr}\sigma_{od}=\Phi_{odr}\sigma_{od},
\end{equation}
by using $\sigma_{od}^2=1$ and defining a local order parameter for triad
balance, $\Phi_{odr} \equiv \sigma_{od}\sigma_{or}\sigma_{dr}$.
The stationarity condition requires $\sigma_{od}' = \Phi_{odr} \sigma_{od} = \sigma_{od}$
for an arbitrary triangle of $o$, $d$, and $r$. We thus conclude that stationarity is equivalent to the balance condition that $\Phi_{odr}=1$~\cite{oishi2021balanced}.
As for L4, recall that its only difference from L6 is that a good donor's cooperation toward a bad recipient is regarded as good. However, the difference will never be observed in a balanced configuration if everyone follows the L4 rule, which prescribes a good donor to defect against a bad recipient. This indicates that balance implies stationarity in L4.
The converse can be proved by enumerating all the possible cases of $(\sigma_{or}, \sigma_{dr}, \sigma_{od})=(\pm 1, \pm 1, \pm 1)$ and checking whether $\sigma_{od}'=\sigma_{od}$ regardless of the order of sampling the three players.

Let us imagine all the possible $2^{N\times N}$ edge configurations on a complete graph, only one of which is the paradise. The question is whether the paradise will occupy $100\%$ of the stationary probability in the limit of $\epsilon \to 0$.
To investigate the cluster dynamics under L6, we begin by showing the following: At a balanced configuration with two clusters, a single error can move at most a vertex from a cluster to the other cluster.
Let us start with a balanced configuration dividing $V$ into two partitions, i.e.,
\begin{equation}
P_\text{balance} = \{ \{v_1, \ldots, v_n\}, \{v_{n+1}, \ldots, v_N\}\},
\label{eq:Porig}
\end{equation}
where $1\le n<N$. The paradise is represented by a trivial partition $P_\text{paradise} = \{V\}$, which we regard as a special case of $P_\text{balance}$ with $n=N$ by a slight abuse of notation to allow a partition to be empty.
Without loss of generality, we assume that $v_n$ has committed an assessment error, after which all time evolution strictly follows the common social norm L6.
To be more specific, the focal individual $v_n$ commits an error either by assessing its friend as bad or by assessing its enemy as good, so $V$ can be divided into the following five partitions, some of which may be empty:
\begin{equation}
P_\text{error} = \{ \underbrace{\{\ldots\}}_{\pi_1}, \underbrace{\{ \ldots \}}_{\pi_2}, \{v_n\}, \underbrace{\{ \ldots \}}_{\pi_3}, \underbrace{\{\ldots\}}_{\pi_4}\},
\label{eq:Perr}
\end{equation}
where $\pi_1 \cup \pi_2 = \{v_1, \ldots, v_{n-1}\}$ and $\pi_3 \cup \pi_4 = \{v_{n+1}, \ldots, v_N\}$. The members of $\pi_1$ and $\pi_2$ like $v_n$, but $v_n$ likes $\pi_1$ and dislikes $\pi_2$. The members of $\pi_3$ and $\pi_4$ dislike $v_n$, but $v_n$ likes $\pi_3$ and dislikes $\pi_4$.
All members in each partition are the same in relation to those in another partition, so a partition can behave like a single vertex.
The dynamics can also make the elements of $\pi_1$ go over to $\pi_2$ or vice versa, and the same applies to $\pi_3$ and $\pi_4$, but $v_n$ remains alone in its own partition because the dynamics has no more erroneous assessment.
The important point is that the five-vertex picture in Eq.~\eqref{eq:Perr} will remain valid on the understanding that some of the partitions may be empty.
When we exhaustively trace the time evolution of the corresponding five-vertex configurations by applying L6, the final outcome is always either the original balanced configuration $P_\text{balance}$ [Eq.~\eqref{eq:Porig}] or a new balanced one represented by
\begin{equation}
P'_\text{balance} = \{ \{v_1, \ldots, v_{n-1}\}, \{v_n, \ldots, v_N\}\},
\label{eq:Pnew}
\end{equation}
where $v_n$ now belongs to the latter partition.
In short, an error can cause at most a single vertex to change its membership under L6.

\begin{figure}
    \includegraphics[width=0.8\columnwidth]{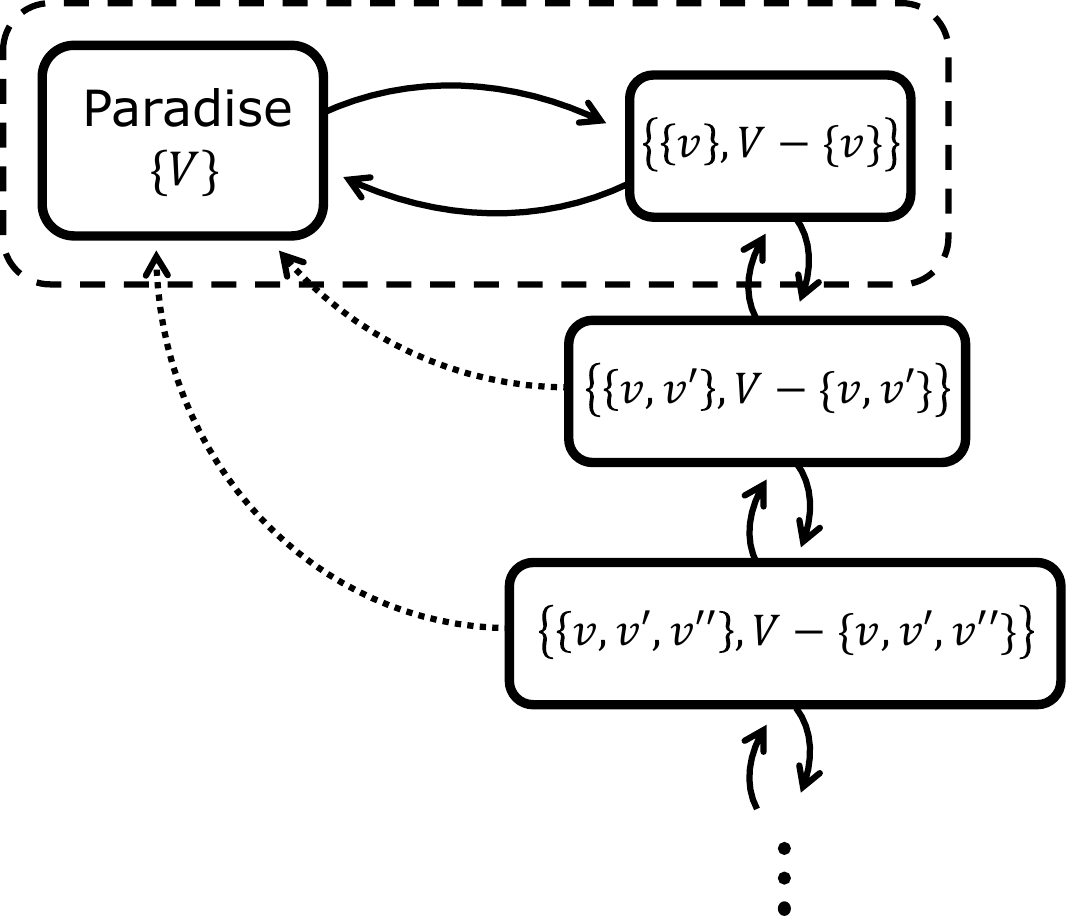}
    \caption{Schematic representation of transitions among balanced configurations under L6 or L4, where $v$, $v'$, and $v''$ are arbitrary elements of $V$. Every arrow in the diagrams denotes a transition induced by a single assessment error, and the dotted arrows exist only in the case of L4.
    The dashed box containing the transition between the paradise $\{V\}$ and $\{\{v\}, V-\{v\}\}$ is detailed in Fig.~\ref{fig:diagram}.
    }
    \label{fig:L4state}
\end{figure}

Let us assume that each player's self-image has converged to $+1$ because $\sigma_{dd}' = \sigma_{dr}^2 = +1$. We may expect this convergence within $O(1)$ Monte Carlo steps~\cite{oishi2021balanced}. In addition, our analysis assumes that no error occurs in self-images.
Then, for any transition sequence of configurations permitted by the above rule of L6, if we choose an arbitrary vertex and flip all its edges, except the self-loop, we will get another valid sequence of configurations permitted by L6.
It can be seen directly from the L6 rule itself: When the chosen vertex is a donor, flipping $\sigma_{od}'$ and $\sigma_{dr}$ makes the equality hold, while leaving $\sigma_{dd}'=+1$ unchanged. When the chosen vertex is a recipient, it assesses the donor with $\sigma_{rd}' = \sigma_{dr}$, whose equality is unaffected by flipping both the sides. Finally, when the chosen vertex is an observer, flipping $\sigma_{od}'$ and $\sigma_{or}$ again satisfies the above rule.
Therefore, for every sequence of transition from Eq.~\eqref{eq:Porig} to Eq.~\eqref{eq:Pnew}, the transformation applied to $v_n$ generates the corresponding sequence in the opposite direction with equal probability.
In addition, as proved above, no more than a single vertex can change its cluster in this $O(\epsilon)$ description, which precludes the existence of any indirect transition paths connecting Eq.~\eqref{eq:Porig} and Eq.~\eqref{eq:Pnew} via a third balanced configuration.
As a consequence, the total transition probability must be the same in either direction, satisfying Kolmogorov's criterion. The resulting detailed-balance condition equalizes the stationary probabilities of Eq.~\eqref{eq:Porig} and Eq.~\eqref{eq:Pnew}.
Extending this argument one by one, we can conclude that every balanced configuration has the same stationary probability.

Let us now consider L4. According to our exhaustive enumeration, not only $P_\text{balance}$ and $P'_\text{balance}$ but also $P_\text{paradise}=\{V\}$ becomes accessible from $P_\text{error}$. Single-error transitions among balanced configurations under L4 are depicted as in Fig.~\ref{fig:L4state}.
The important point is the existence of the one-way transition to the paradise, when both of the clusters have more than one vertex. Such balanced configurations should have only negligible stationary probabilities because of the one-way transition.
Still, the stationary probability of the paradise does not necessarily approach
$100\%$ even with small $\epsilon$ because the system may go back and forth
between the paradise and $\{\{v\}, V-\{v\}\}$. For example, if $N=4$, our
numerical estimate of the stationary distribution shows that
$\rho_\text{st} \left[ \left\{V\right\} \right]$ occupies only $61\%$ when $\epsilon = 10^{-4}$, which is consistent with the order-counting argument~\cite{murase2020five}.

\begin{widetext}
\begin{figure*}[t]
    \includegraphics[width=0.7\textwidth]{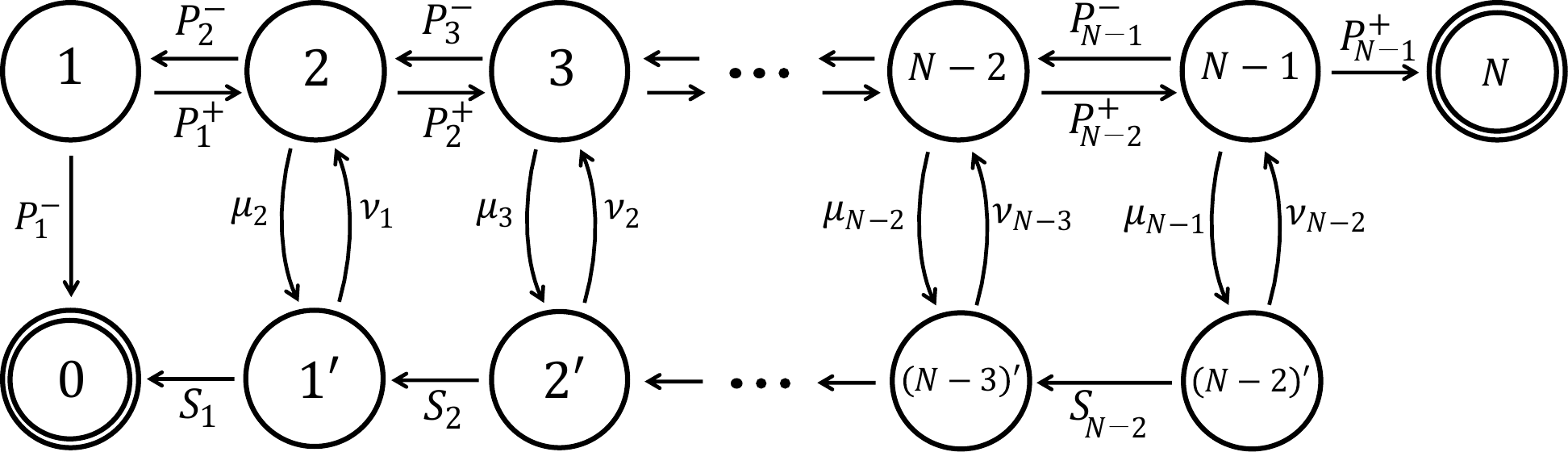}
    \caption{Zoomed-in views of the dashed box in Fig.~\ref{fig:L4state}, containing the transition between the paradise and $\{\{v\}, V-\{v\}\}$. Self-loops are all omitted for the sake of brevity. Each node represents a set of configurations that are identical up to the permutation of vertices.
    The paradise and $\{\{v\}, V-\{v\}\}$ are labeled with $0$ and $N$, respectively.
    The symbols attached to edges are transition probabilities:
    We have $P_j^+ = (N-j)/N^2$,
    whereas $P_j^- = (N-j)(j-1)/N^2$ for $j>1$ with $P_1^- = (N-1)/N^2$.
    The other transition probabilities are $\mu_j = (N-j)/N^2$, $\nu_{j} = j/N^2$, and $S_j = j(N-j)/N^2$.
    }
    \label{fig:diagram}
\end{figure*}
\end{widetext}

However, the transition probability from $\{\{v\}, V-\{v\}\}$ to the paradise becomes far greater than the opposite one as $N$ grows. Figure~\ref{fig:diagram} provides a zoomed-in view for the dashed box of Fig.~\ref{fig:L4state}, where the nodes denoted by $0$ and $N$ correspond to the paradise and $\{\{v\}, V-\{v\}\}$, respectively. Each node represents a set of edge configurations that are identical up to permutation of vertices.
The absorption probability into $\{\{v\}, V-\{v\}\}$ from configuration $j$ is denoted by $q_j$, which means that $q_0=0$ and $q_N=1$ by definition. The symbols attached to the arrows in Fig.~\ref{fig:diagram} are transition probabilities.

If an assessment error occurs at the paradise, the starting point of the transition dynamics will be the configuration denoted by $1'$. The dynamics will end at $\{\{v\}, V-\{v\}\}$ with probability $q_{1'}$. Likewise, if an error occurs at $\{\{v\}, V-\{v\}\}$ by assessing an enemy as good,
the dynamics starts from the node denoted by $N-1$, from which the probability of absorption into the paradise is $1-q_{N-1}$.
We compare $q_{1'}$ with $1-q_{N-1}$ to see the asymmetry in the probability flows:
The solution in the Supplemental Material~\cite{suppl} shows that $(1-q_{N-1}) / q_{1'}$ is a rapidly increasing function of $N$, which outputs $15$, $82$, and $517$ for $N=4$, $5$, and $6$, respectively.
It is instructive to consider the random walk only among the upper nodes without primes to get a lower bound of such asymmetry because the existence of the lower nodes, labeled with primes, should bias the net probability flow even more drastically toward the paradise.
Although $q_{1'}$ does not appear in this simplified problem, one can readily see that it is proportional to $q_1$.
By solving a one-dimensional random walk with position-dependent transition probabilities~\cite{nowak2006evolutionary}, we can readily see $(1-q_{N-1})/q_1 = (N-2)!$, which confirms that the net probability flow becomes more and more biased toward the paradise as $N$ increases.
One could point out that we have not taken into account the existence of other paths from $\{\{v\}, V-\{v\}\}$ to the paradise, e.g., through an error inside the larger cluster, but they can only contribute positively to the bias. We discuss more details on the large-$N$ behavior under L4 in the Supplemental Material~\cite{suppl}.

\begin{figure}
    \includegraphics[width=0.45\columnwidth]{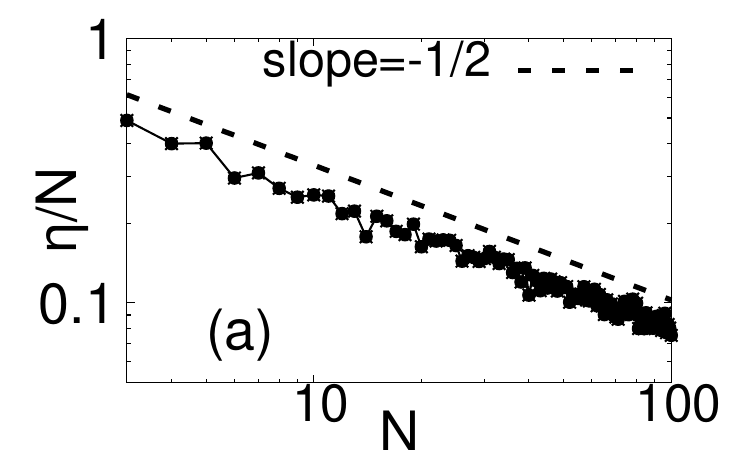}
    \includegraphics[width=0.45\columnwidth]{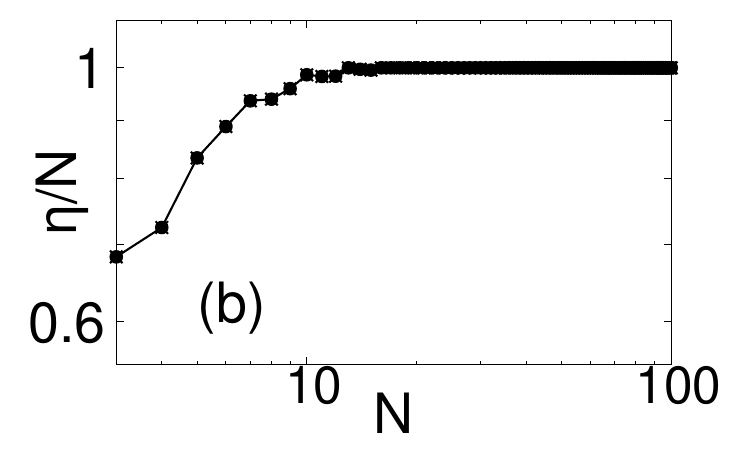}
    \caption{Absolute difference of cluster sizes divided by $N$, obtained from $2\times 10^2$ Monte Carlo samples. Each sample starts from a random initial configuration and follows the common social norm with $\epsilon=0$ until arriving at a balanced configuration.
    We have defined $\eta \equiv \left| n - (N-n) \right|$, assuming that the clusters are of sizes $n$ and $N-n$, respectively. (a) Under L6, the behavior of $\eta/N$ is proportional to $N^{-1/2}$, which is fully consistent with a prediction from the binomial distribution. (b) Under L4, by contrast, $\eta/N$ tends to $1$ as $N$ grows, which implies the emergence of the paradise. The error bars are smaller than the symbol size.}
    \label{fig:diff}
\end{figure}

The above analysis shows that the segregation phenomenon induced by L6 is driven by entropy in the sense that the number of segregated balanced configurations greatly exceeds the unique possibility of the paradise as $N$ grows.
As long as every balanced configuration has equal probability,
the most probable case would be such that two clusters are of roughly equal sizes as in the binomial distribution.
We can demonstrate it by using Monte Carlo simulations [Fig.~\ref{fig:diff}(a)], according to which the size difference between two clusters, divided by $N$, decreases as $N^{-1/2}$.
Note the different behavior of L4, according to which the paradise is easily accessible when $N \gtrsim O(10)$ [Fig.~\ref{fig:diff}(b)].
This prediction of L6 is also partially supported by a recent analytic study~\cite{fujimoto2022reputation}, showing that each individual is expected to receive good assessments from half of the population, although it does not distinguish a clustered configuration from a randomly mixed one.

It is worth pointing out that L6 is a remarkably simple and intuitive norm as
indicated by its nickname ``stern judging'': It assesses cooperation (defection) toward a good person as good (bad), and cooperation (defection) toward a bad person as bad (good). It thus prescribes cooperation toward a good person and defection toward a bad person. Despite all its good properties, L6 divides a fully connected society into two mutually hostile clusters of roughly equal sizes.
The segregation might seem to be an ordered configuration compared to a random
mixture, but the dominant factor behind it turns out to be entropy, and who belongs to which side is only a matter of chance.
We propose L4 as a remedy for those problems: This norm leads a sufficiently large society of $N \gtrsim O(10)$ to the paradise just by changing the assessment of a good person's cooperation toward a bad person.
To see when cooperation is secured by this remedy, we discuss
evolutionary aspects in the Supplemental Material~\cite{suppl}.

From a broader perspective, the precise understanding of segregation on fully
connected graphs may give us theoretical insights into the recent trend of
increasing political polarization in our hyperconnected society, as well as how
to mitigate the trend with a minimal change in our judging behavior. Considering
that the basic dynamics of a social norm would be preserved in a more realistic acquaintance network structure~\cite{kuroda2023frustrated}, we expect that our findings should also be relevant to real social phenomena.

\section*{}
We gratefully acknowledge helpful comments from Yohsuke Murase.
M.B. and S.K.B. acknowledge support by Basic Science Research Program through the
National Research Foundation of Korea (NRF) funded by the Ministry of Education
(NRF-2020R1I1A2071670).
This work was supported by the Global Joint Research Program funded by the Pukyong National University (202411500001).
We appreciate the APCTP for its hospitality during the completion of this work.

\onecolumngrid
\appendix

\section{Balance and stationarity}
\label{app:balance}

\subsection{L6 and L4}

We may rewrite the L6 rule as
\begin{equation}
\sigma_{od}'=\sigma_{or}\sigma_{dr}=\sigma_{od}\sigma_{or}\sigma_{dr}\sigma_{od}=\Phi_{odr}\sigma_{od},
\end{equation}
by using $\sigma_{od}^2=1$ and defining a local order parameter for triad balance,
\begin{equation}
    \Phi_{odr} \equiv \sigma_{od}\sigma_{or}\sigma_{dr}.
\end{equation}
The stationarity condition requires $\sigma_{od}' = \Phi_{odr} \sigma_{od} = \sigma_{od}$
for an arbitrary triangle of $o$, $d$, and $r$. We thus conclude that stationarity is equivalent to the balance condition that $\Phi_{odr}=1$~\cite{oishi2021balanced}.

The same statement holds true for L4. The assessment rule of L4 can be written as follows~\cite{lee2022second}:
\begin{equation}
    \sigma_{od}'=\frac{1}{4}\left(-\sigma_{or}+3\sigma_{dr}\sigma_{or}+\sigma_{dr}+\sigma_{dr}\sigma_{od}-\sigma_{or}\sigma_{od}+1+\sigma_{od}-\sigma_{dr}\sigma_{or}\sigma_{od} \right).
    \label{eq:L4}
\end{equation}
If the balance condition is met, we have $\Phi_{odr}=\sigma_{od}\sigma_{or}\sigma_{dr}=1$. It automatically means
that
\begin{subequations}
\begin{align}
\sigma_{od} &= \sigma_{or}\sigma_{dr}\\
\sigma_{or} &= \sigma_{od}\sigma_{dr}\\
\sigma_{dr} &= \sigma_{od}\sigma_{or}
\end{align}
\label{eq:corollary}
\end{subequations}
because $\sigma_{od}^2 = \sigma_{or}^2 = \sigma_{dr}^2 = 1$. Plugging these relations into Eq.~\eqref{eq:L4}, we obtain $\sigma_{od}' = \sigma_{od}$ for an arbitrary $od$-pair, which means stationarity.
Now, we have to check if the converse is true. The stationarity condition requires
$\sigma_{od}' = \sigma_{od}$, which leads to
\begin{equation}
    \sigma_{od} = \frac{1}{4} \left( -\sigma_{or}+3\sigma_{dr}\sigma_{or}+\sigma_{dr}+\sigma_{dr}\sigma_{od}-\sigma_{or}\sigma_{od}+1+\sigma_{od}-\sigma_{dr}\sigma_{or}\sigma_{od} \right).
    \label{eq:L4station}
\end{equation}
If we enumerate the eight possible cases of $(\sigma_{or}, \sigma_{dr}, \sigma_{od})=(\pm 1, \pm 1, \pm 1)$, Eq.~\eqref{eq:L4station} is satisfied for $(+1,+1,+1)$, $(-1,-1,+1)$, $(-1,+1,-1)$, $(+1,-1,-1)$, and $(-1,+1,+1)$, among which only the last one is unbalanced. To eliminate the last possibility, we note that stationarity is actually a stronger condition than Eq.~\eqref{eq:L4station}: For a given configuration to be stationary, it must be left invariant even if we sample the three players in different order, so that $(\sigma_{or}, \sigma_{dr}, \sigma_{od})=(+1,-1,+1)$ and $(+1,+1,-1)$ must also satisfy Eq.~\eqref{eq:L4station}, while neither of them is in the list.
We thus exclude $(-1,+1,+1)$ from consideration and conclude that the balance condition is equivalent to stationarity in L4.

\subsection{L3, L5, L7, and L8}

By contrast, the balance condition is not equivalent to stationarity in L3 and L5. The assessment rule of L3 can be expressed as follows:
\begin{equation}
    \sigma_{od}'=\frac{1}{2} \left( 1+ \sigma_{dr} -\sigma_{or} +\sigma_{or}\sigma_{dr} \right).
    \label{eq:L3}
\end{equation}
Suppose that the balance condition $\Phi_{odr}=\sigma_{od}\sigma_{or}\sigma_{dr}=1$ is met. By using Eq.~\eqref{eq:corollary}, we may rewrite Eq.~\eqref{eq:L3} as
\begin{equation}
    \sigma_{od}'=
    \frac{1}{2} \big[ 1+\sigma_{od}(1+\sigma_{or}-\sigma_{dr}) \big].
\end{equation}
If $(\sigma_{or}, \sigma_{dr}, \sigma_{od})=(-1,+1,-1)$, we see that stationarity is violated because $\sigma_{od}' \neq \sigma_{od}$.

Likewise, the assessment rule of L5 can be represented as follows:
 \begin{equation}
    \sigma_{od}'=\frac{1}{4} \left( -\sigma_{or}+3\sigma_{dr}\sigma_{or}+\sigma_{od}\sigma_{or}+1-\sigma_{od}+\sigma_{dr}\sigma_{od}\sigma_{or}+\sigma_{dr}-\sigma_{dr}\sigma_{od} \right)
 \end{equation}
Under the balance condition, together with Eq.~\eqref{eq:corollary}, we obtain
\begin{eqnarray}
    \sigma_{od}'&=&\frac{1}{4}[-\sigma_{or}+3\sigma_{dr}\sigma_{or}+\sigma_{od}\sigma_{or}+1-\sigma_{od}+1+\sigma_{dr}-\sigma_{dr}\sigma_{od}] \\
    &=& \frac{1}{4}[-2\sigma_{or}+2\sigma_{od}+2\sigma_{dr}+2] = \frac{1}{2}[1+\sigma_{od}-\sigma_{or}+\sigma_{dr}]
\end{eqnarray}
Again, stationarity is violated if $(\sigma_{or}, \sigma_{dr}, \sigma_{od})=(-1,+1,-1)$. The balance condition is not equivalent to stationarity.

For L7 and L8, it is straightforward to prove inequivalence between stationarity and the balance condition. The reason is that both assign bad reputation when a bad donor defects against a bad recipient. A triad of $(\sigma_{or}, \sigma_{dr}, \sigma_{od})=(-1,-1,-1)$ is thus an absorbing configuration, although it is not balanced.

\subsection{L1 and L2}

For L1 and L2, a donor's action to a recipient cannot simply be identified with $\sigma_{dr}$ from the beginning because a bad donor should cooperate to a bad recipient, i.e., $\beta_{BB}=C$. For L1, a donor's action can be coded as
\begin{equation}
    \beta_d (\sigma_{dd}, \sigma_{dr}) = \frac{1}{2} \left[ 1+\sigma_{dd} (\sigma_{dr}-1) + \sigma_{dr} \right],
    \label{eq:L1action}
\end{equation}
and an observer's assessment rule is given as
\begin{equation}
    \sigma_{od}' = \frac{1}{4} \left[ 1+\sigma_{od}+3\beta_d - \sigma_{od}\beta_d + (1+\sigma_{od})(\beta_d-1)\sigma_{or} \right].
    \label{eq:L1assessment}
\end{equation}
When $o=d$, we obtain $\sigma_{dd}'=+1$ regardless of $\sigma_{dd}$ and $\sigma_{dr}$, so each donor's self-evaluation quickly converges to $+1$. If $\sigma_{dd}=+1$ is plugged into Eq.~\eqref{eq:L1action}, we can identify $\beta_d$ with $\sigma_{dr}$ as in the previous cases.
Similarly, the assessment rule of L2 is
\begin{equation}
    \sigma_{od}' = \frac{1}{2} \beta_d \left[ 1+\sigma_{od} (\sigma_{or}-1) + \sigma_{or} \right],
\end{equation}
where $\beta_d$ is given by Eq.~\eqref{eq:L1action}. By setting $o=d$, we once again see the convergence toward $\sigma_{dd}'=+1$ regardless of $\sigma_{dd}$ and $\sigma_{dr}$, which allows us to identify $\beta_d$ with $\sigma_{dr}$.

Therefore, for both L1 and L2, as soon as a donor's action is fully prescribed from $\sigma_{dr}$, we can apply the same argument as above to prove the inequivalence between stationarity and the balance condition: A bad donor's defection against a bad recipient is judged as bad by these norms, and this shows why $(\sigma_{or}, \sigma_{dr}, \sigma_{od})=(-1,-1,-1)$ is an unbalanced absorbing configuration.

\section{Number of balanced states}
\label{app:induction}

Assume that we have a set of vertices $V = \{v_1, \ldots, v_N\}$.
Thanks to the structure theorem, we just have to find the number of ways to divide those $N$ elements into two clusters. If none of the clusters is empty, we can prove that the answer is
\begin{equation}
     W_N = 2^{N-1}-1
     \label{eq:induction}
\end{equation}
by using mathematical induction. First, if $N=1$, we have no way to divide it into two nonempty clusters, which means that $W_1 = 0$. Second, let us assume that
\begin{equation}
    W_k = 2^{k-1}-1
    \label{eq:assumption}
\end{equation}
for some positive integer $k$. When a new vertex $v_{k+1}$ has appeared, we have two possibilities to divide the $k+1$ elements into two clusters. One is to add $v_{k+1}$ to one of the existing clusters. The other is to make a single-element cluster of $v_{k+1}$ and merge the existing clusters into one. In other words, we have
\begin{equation}
    W_{k+1} = 2 W_{k} + 1.
\end{equation}
Plugging Eq.~\eqref{eq:assumption} here, we find that
\begin{equation}
    W_{k+1} = 2 \left( 2^{k-1}-1 \right) +1 = 2^{k}-1,
\end{equation}
which proves Eq.~\eqref{eq:induction} for general $N$. Note that $W_N$ does not include the paradise. If we take it into account, the number of balanced states is $B = 2^{N-1}$.

\setcounter{figure}{0}
\renewcommand{\thefigure}{C\arabic{figure}}

\section{Recurrence formulas for L4}
\label{app:recurL4}

\begin{figure}
    \includegraphics[width=0.95\columnwidth]{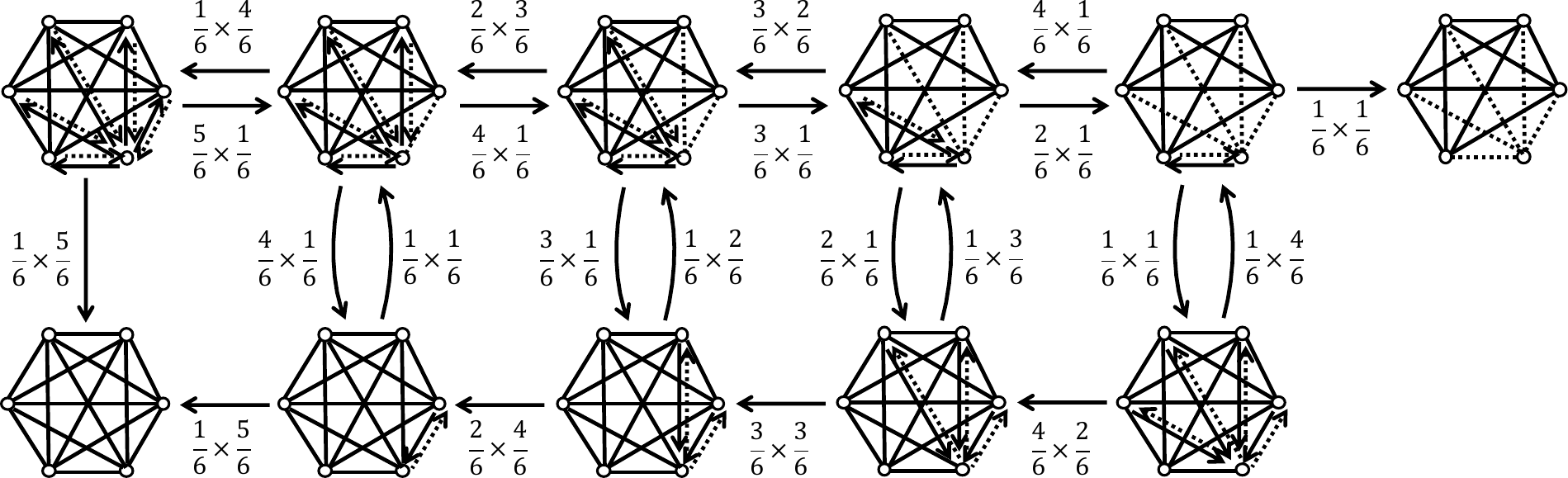}
    \caption{
    An example of Fig.~2 when $N=6$. The lower left corner is the paradise,
    denoted by $0$ in Fig.~2, whereas the upper right corner shows the
    segregation of a single individual from the others, denoted by $1$ in Fig.~2.}
    \label{fig:diagram_suppl}
\end{figure}

For the sake of concreteness, Fig.~\ref{fig:diagram_suppl} shows an example of
Fig.~2 when $N=6$.
To solve the problem through recurrence formulas, let us write equations for the absorption probabilities as follows:
\begin{subequations}
\begin{align}
q_j &= P_j^- q_{j-1} + P_j^+ q_{j+1} + \mu_j q_{(j-1)'} + \left(1-P_j^- -P_j^+ -\mu_j \right) q_j\\
q_{k'} &= S_k q_{(k-1)'} + \nu_{k} q_{k+1} + \left(1-S_k-\nu_{k} \right) q_{k'},
\end{align}
\end{subequations}
where $j=1,\ldots,N-1$ and $k=1,\ldots,N-2$. By definition, we have $q_0 \equiv 0$ and $q_N=1$. It is also convenient to define $q_{0'} \equiv 0$ and $\mu_1 \equiv 0$. We may rewrite the above formulas as
\begin{subequations}
\begin{align}
0 &= P_j^- \left(q_{j-1} - q_j\right) + P_j^+ \left(q_{j+1} - q_j \right) + \mu_j \left(q_{(j-1)'} - q_j \right)\label{eq:recur1}\\
q_{k+1} &= \Gamma_k \left( q_{k'} - q_{(k-1)'} \right) + q_{k'},
\label{eq:qj}
\end{align}
\end{subequations}
where $\Gamma_k \equiv S_k / \nu_{k}$.
Let us define $R_k \equiv q_{k'} - q_{(k-1)'}$. Plugging Eq.~\eqref{eq:qj} into Eq.~\eqref{eq:recur1} and rearranging the terms, we obtain
\begin{equation}
P_j^-\Gamma_{j-2}R_{j-2}-\left[P_j^-(\Gamma_{j-1}+1)+P_j^+\Gamma_{j-1}+\mu_j\Gamma_{j-1}\right]R_{j-1}\\
+P_j^+(\Gamma_j +1)R_j=0.
\label{eq:recur2}
\end{equation}

We now need to write transition probabilities.
In Fig.~2, the configuration denoted by $1$ means that a vertex $v$ is disliked by all others whereas it still likes them. The next one denoted by $2$ means that $v$ now dislikes one of them. The transition from $1$ to $2$ occurs with probability $P_1^+ = 1/N \times (N-1)/N$ because $v$ must be the recipient with probability $1/N$ while anyone else can be chosen as the donor with probability $(N-1)/N$.
The reverse transition from $2$ to $1$ happens with probability $P_2^- = (1/N) \times (N-2)/N$ because the one who $v$ dislikes ($1$ out of $N$) must donate someone who $v$ still likes ($N-2$ out of $N$). In general, we should have $P_j^- = (j-1)/N \times (N-j)/N$. The exception is $P_1^- = 1/N\times (N-1)/N$ because the corresponding transition occurs when $v$ donates to anyone else.
The transition probabilities between unprimed nodes are thus given as follows:
\begin{subequations}
\begin{align}
P_j^+ &= \left(\frac{N-j}{N}\right) \left(\frac{1}{N}\right)\\
P_j^- &= \left\{
\begin{array}{ll}
\left(\frac{N-1}{N}\right) \left(\frac{1}{N}\right) & \text{for~}j=1\\
\left(\frac{j-1}{N}\right) \left(\frac{N-j}{N}\right) & \text{for~}j>1,
\end{array}\right.
\end{align}
\end{subequations}

The transition probability $\mu_j$ corresponds to the case of choosing  $v$ as the donor ($1$ out of $N$) and someone else who $v$ likes as the recipient ($N-j$ out of $N$). Its counterpart $\nu_{j}$ is the probability of choosing $v$ as the donor ($1$ out of $N$) and someone else who $v$ dislikes as the recipient ($j$ out of $N$). Finally, $S_j$ is the probability of choosing someone who $v$ dislikes as the donor ($j$ out of $N$) and someone else who $v$ likes (it can be $v$ itself) as the recipient ($N-j$ out of $N$). Therefore, their general expressions are
\begin{subequations}
\begin{align}
\mu_j &= \left(\frac{N-j}{N}\right) \left(\frac{1}{N}\right)\\
\nu_{j} &= \left(\frac{1}{N}\right) \left(\frac{j}{N}\right)\\
S_j &= \left(\frac{j}{N}\right) \left(\frac{N-j}{N}\right),
\end{align}
\end{subequations}
which results in $\Gamma_j = S_j / \nu_{j} = N-j$.

At $j=1$, Eq.~\eqref{eq:recur2} reduces to
\begin{equation}
P_1^- \left(-q_1\right) + P_1^+ \left( \Gamma_1 q_{1'} + q_{1'} - q_1 \right) = 0,
\end{equation}
from which we obtain
\begin{equation}
R_1 = q_{1'} = \frac{P_1^+ + P_1^-}{P_1^+ \left(\Gamma_1+1\right)} q_1 = \frac{2}{N} q_1.
\label{eq:q1'}
\end{equation}
At $j=2$, Eq.~\eqref{eq:recur2} takes the following form:
\begin{equation}
P_2^- \left[q_1 - \left(\Gamma_1 +1 \right)R_1 \right] + P_2^+ \left[ \left(\Gamma_2+1\right) R_2 - \Gamma_1 R_1\right] - \mu_2 \Gamma_1 R_1 = 0,
\end{equation}
from which we find
\begin{equation}
R_2 = \frac{5N-4}{N(N-1)}q_1.
\end{equation}
From $j=3$ to $N-2$, we can find every $R_j$ from the following recurrence formula:
\begin{equation}
R_j = \frac{j-j^2+N+jN}{N-j+1} R_{j-1} - \frac{(j-1)(N-j+2)}{N-j+1} R_{j-2}.
\end{equation}
Although we have $R_k$'s with $k=1,\ldots,N-2$, the overall factor of $q_1$ remains unknown, and it has to be determined from Eq.~\eqref{eq:recur2} at $j=N-1$:
\begin{eqnarray}
&&P_{N-1}^- \left[ \Gamma_{N-3} R_{N-3} - \left(1+\Gamma_{N-2} \right) R_{N-2} \right]\nonumber\\
&+&P_{N-1}^+ \left[ \left(1-q_{(N-2)'}\right)- \Gamma_{N-2} R_{N-2} \right] - \mu_{N-1} \Gamma_{N-2} R_{N-2} = 0,
\label{eq:j=N-1}
\end{eqnarray}
where we have used $q_N=1$. By using the following identity,
\[q_{(N-2)'} = \left(q_{(N-2)'} - q_{(N-3)'}\right) +
\left(q_{(N-3)'} - q_{(N-4)'}\right) + \ldots +
\left(q_{2'} - q_{1'}\right) +
\left(q_{1'} - q_{0'}\right)
= \sum_{k=1}^{N-2} R_k,\]
we can write Eq.~\eqref{eq:j=N-1} in terms of $R_j$ only:
\begin{equation}
-(3N-6) R_{N-3} + (3N-2) R_{N-2} + \sum_{k=1}^{N-2} R_k = 1,
\end{equation}
from which we determine the value of $q_1$.

\setcounter{figure}{0}
\renewcommand{\thefigure}{D\arabic{figure}}

\section{Dominance of the paradise under L4 when $N \to \infty$}
\label{app:direct}

In the main text, we have shown that one of the transitions shown in Fig.~1 becomes negligible for sufficiently large $N$. Then, one might ask whether other transitions survive the same large $N$. This is a relevant question because the stationary probability distribution might be affected if some other transitions also disappear: For example, the paradise will fail to occupy 100\% if we have a set of segregated configurations, from which neither the one-way transition to the paradise nor the transition to $\{\{v\}, V-\{v\}\}$ occurs. For this reason, here we wish to argue that L4 will surely arrive at the paradise even in the large-$N$ limit.

Let us divide all the balanced configurations into three categories: The paradise, single-enemy configurations represented by $\left\{ \{v\}, V - \{v\} \right\}$, and all the others. Let their probabilities be denoted as $\rho_0, \rho_1$, and $\rho_2$, respectively. Then, their transition dynamics is written as
\begin{equation}
    \begin{pmatrix}
        \rho_0 \\ \rho_1 \\ \rho_2
    \end{pmatrix}'
    =
    \begin{pmatrix}
        1-\epsilon q_{1'} & \frac{\epsilon(1-{q_{N-1}})}{N} & \epsilon P_{2 \to 0} \\
        \epsilon q_{1'} & 1-\frac{\epsilon(1-q_{N-1})}{N} - \epsilon P_{1 \to 2} & \epsilon P_{2 \to 1} \\
        0 & \epsilon P_{1 \to 2} & 1- \epsilon P_{2 \to 0} - \epsilon P_{2 \to 1}
    \end{pmatrix}
    \begin{pmatrix}
        \rho_0 \\ \rho_1 \\ \rho_2
    \end{pmatrix},
\end{equation}
where $P_{n \to n'}$ is the probability to reach a configuration classified as
$n'$, given an error at $n$. Now we estimate $P_{1\to 2}$ from Fig.~2:
If we start from the segregated state on the rightmost side, the probability to transit to the next node denoted by $N-1$ is $\epsilon/N$ because we have to choose one of $N-1$ negative edges out of $N(N-1)$ edges.
Therefore, under the condition that an error has occurred, the probability to reach a third balanced configuration other than the paradise and $\{\{v\}, V-\{v\}\}$ is bounded from above by $1-1/N$.
For this reason, we take the worst-case scenario that $P_{1 \to 2} = 1-1/N$ and $P_{2 \to 1} = 0$, which leads to the following transition dynamics:
\begin{equation}
    \begin{pmatrix}
        \rho_0 \\ \rho_1 \\ \rho_2
    \end{pmatrix}'
    =
    \begin{pmatrix}
        1-\epsilon q_{1'} & \frac{\epsilon(1-{q_{N-1}})}{N} & \epsilon P_{2 \to 0} \\
        \epsilon q_{1'} & 1-\epsilon+\frac{\epsilon q_{N-1}}{N} & 0 \\
        0 & \epsilon\left(1- \frac{1}{N}\right) & 1- \epsilon P_{2 \to 0}
    \end{pmatrix}
    \begin{pmatrix}
        \rho_0 \\ \rho_1 \\ \rho_2
    \end{pmatrix}.
\end{equation}
The principal eigenvector is readily obtained, and it takes the following form before normalization in the limit of $N\to \infty$:
\begin{equation}
    \begin{pmatrix}
        \rho_0^* \\ \rho_1^* \\ \rho_2^*
    \end{pmatrix}
    \xrightarrow[N\to \infty]{}
    \begin{pmatrix}
        P_{2 \to 0}/ q_{1'} \\
        P_{2 \to 0} \\
        1
    \end{pmatrix}.
    \label{eq:principal}
\end{equation}
If we define
\begin{equation}
\gamma_j \equiv P_j^- / P_j^+ =
\left\{ \begin{array}{ll}
1 & \text{for~}j=1\\
j-1 & \text{for~}j>1,
\end{array}
\right.
\end{equation}
it is straightforward in the one-dimensional random-walk problem to find that
\begin{subequations}
\begin{align}
q_1 &= \left[1+\sum_{i=1}^{N-1} \left( \prod_{j=1}^i \gamma_j \right)\right]^{-1}\label{eq:q1}\\
q_{k+1} &= \left[ 1+\sum_{i=1}^{k} \left( \prod_{j=1}^i \gamma_j \right) \right] q_1,
\end{align}
\end{subequations}
where $k=1,\ldots,N-2$.
Equation~\eqref{eq:q1} shows that
\begin{equation}
    q_1^{-1} = \left[1+\sum_{i=1}^{N-1} \left( \prod_{j=1}^i \gamma_j \right)\right] \ge \left[1+\left( \prod_{j=1}^{N-1} \gamma_j \right)\right] = 1+ (N-2)!,
\end{equation}
and we have $q_{1'}^{-1} = (N/2) q_1^{-1} \sim (N-1)!$ [Eq.~\eqref{eq:q1'}].

Before proceeding, we note that assuming $P_{2\to 1}=0$ must have been a gross oversimplification if we consider the following points by comparing L4 with L6: The number of positive edges is minimized (maximized) when a balanced configuration has two equally large clusters (a single cluster). We can say that L4 tends to have more positive edges than L6 because $\alpha_{GCB}=G$ is its sole difference from L6. It implies the existence of non-negligible probability flow from $\{\{v_1, \ldots, v_n\}, V-\{v_1, \ldots, v_n\}\}$ to $\{\{v\}, V-\{v\}\}$, where $n \ge 2$, under the rule of L4.

\begin{figure}
    \includegraphics[width=0.5\textwidth]{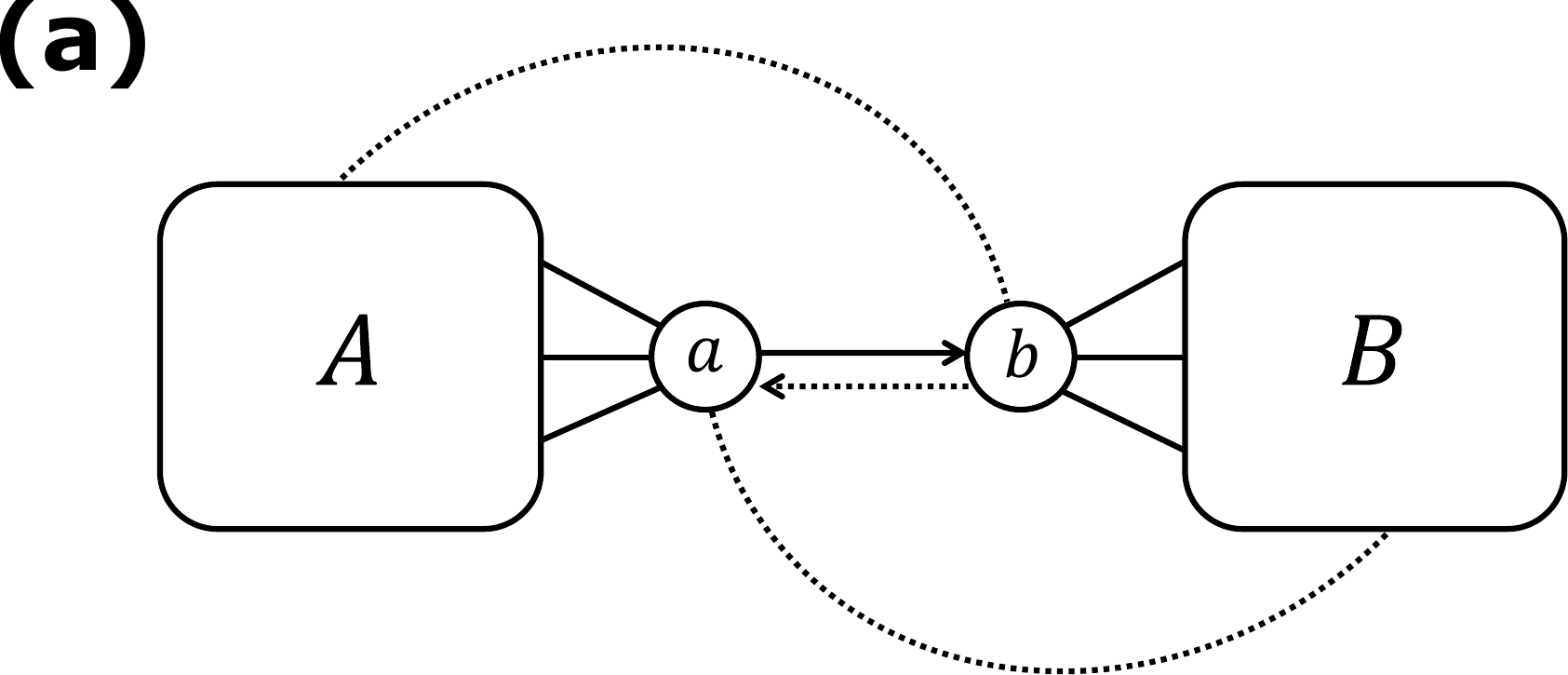}
    \includegraphics[width=0.5\textwidth]{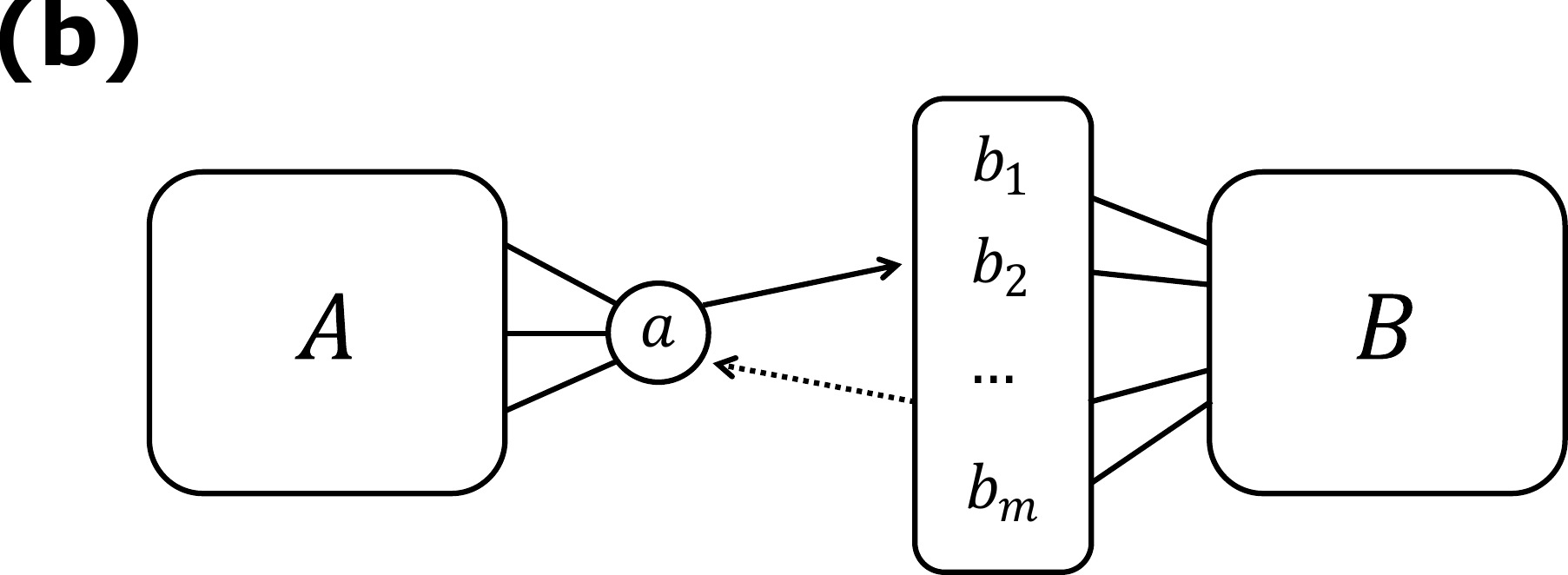}
    \caption{Schematic diagrams to specify trajectories from a segregated configuration to the paradise. (a) The initial segregation between $A \cup \{a\}$ and $B \cup \{b\}$ is perturbed by $a$'s misjudgment of $b$. (b) If a member of $B$ donates to $b$, he or she will also be regarded as good by $a$. In this way, $a$ can assess $b_1, b_2, \ldots, b_m$ as good. This configuration can change to the paradise as soon as $a$ is chosen as the donor and one of $b_i$’s is chosen as the recipient.}
    \label{fig:AabB}
\end{figure}

If we accept the assumption for the worst-case scenario, the problem boils down to the behavior of $P_{2\to 0}$ in the large-$N$ limit. By definition, we can think of $P_{2\to 0}$ as the following weighted average:
\begin{equation}
    P_{2\to 0} = \frac{\sum_{N_1 = 2} T(\text{paradise} | N_1) \text{Pr}(N_1) }{\sum_{N_1 = 2}\text{Pr}(N_1)},
    \label{eq:P_2to0}
\end{equation}
where $\text{Pr}(N_1)$ means the probability to find a configuration of two clusters with sizes $N_1$ and $N_2 = N-N_1$, respectively, and we may assume $N_1 > N_2$ without loss of generality. By $T(\text{paradise} | N_1)$, we mean the transition probability from the configuration to the paradise. If we consider Eqs.~\eqref{eq:principal} and \eqref{eq:P_2to0}, one possible scenario is that $P_{2 \to 0}$ decays slowly compared with $q_{1'}$ so that the paradise appears as the only attractor, i.e., $\rho_0^\ast \to 1$ as $N \to \infty$.
Another possible scenario is that $P_{2 \to 0}$ vanishes, with $P_{2 \to 0}/q_{1'}$ converging to a finite value, which means that the basin of attraction for the paradise is limited to its nearby balanced configurations, so that $\rho_2^\ast$ can be comparable with $\rho_0^\ast$.

We will argue that the first ``paradise'' scenario is the case, by considering a set of specific trajectories from segregation to the paradise, which gives a lower bound of $T(\text{paradise}|N_1)$.
Assume that we choose a donor from the first cluster and its recipient from the other. The probability of such a choice is
\begin{equation}
    \frac{N_1 N_2}{N(N-1)} \approx \phi_1\phi_2,
    \label{eq:rho1_rho2}
\end{equation}
where $\phi_1 \equiv N_1/N$ and $\phi_2 \equiv N_2/N$.
To escape from this balanced configuration, we assume that $a$ erroneously judges $b$ as good. The resulting `excited state' has been depicted in Fig.~\ref{fig:AabB}(a), where the numbers of vertices in $A$ and $B$ are $N_1-1$ and $N_2 -1$, respectively. Let this configuration be denoted by $| 1 \rangle$.
We then imagine that a member of $B$ donates to $b$, by which $a$ will assess the member as good. The resulting configuration may be denoted by $|2\rangle$.
In this way, starting from $|1\rangle$, we can generate $|m\rangle$ in which $a$ unilaterally likes $b_1, b_2, \dots, b_m$, while the other nodes have the same relations with each other as in the initial configuration [Fig.~\ref{fig:AabB}(b)]. Let us estimate the corresponding probability as follows: When we choose a donor $d$ and a recipient $r$ in $|m\rangle$, the following four cases leave the configuration unaltered:
\begin{itemize}
\item $d \in A \cup \{a\}$ and $r \in A \cup \{a\}$
\item $d \in B$ and $r \in B$
\item $d \in A \cup \{a\}$ and $r \in B$
\item $d \in B$ and $r \in A$
\end{itemize}
Therefore, if $X$ is the number of edges between $d$ and $r$ to change the configuration, it is obtained as
\begin{equation}
\begin{split}
    X&\equiv \left[N(N-1)-N_1(N_1-1)-(N_2-m)(N_2-m-1) \right.\\
    & \left.-N_1(N_2-m) -(N_1-1)(N_2-m) \right]\\
    &=2m(N-1) + N_2 - m^2 < 3mN.
\end{split}
\end{equation}
If we choose $d$ and $r$ from $B$ and $\{b_1, \ldots, b_m\}$, respectively, their interaction raises $| m \rangle$ to $| m+1 \rangle$, and the number of such possibilities is $Z \equiv (N_2-m)m$. Let us define $Y \equiv N(N-1)$ as the total number of edges. Then, the probability of transition from $|m\rangle$ to $|m+1\rangle$ is greater than or equal to
\begin{equation}
    \sum_{k=0}^\infty \left[\left(\frac{Y-X}{Y}\right)^{k} \left(\frac{Z}{Y}\right)\right] = \left(\frac{Z}{Y}\right)\frac{1}{1-\left(1-X/Y\right)}=\frac{Z}{X},
\end{equation}
which is greater than
\begin{equation}
    \frac{(N_2-m)m}{3mN} = \frac{N_2 - m}{3N}.
\end{equation}
As a result, the probability of transition from $|1\rangle $ to $|N_2\rangle$ is greater than
\begin{equation}
    \prod_{m=1}^{N_2-1} \frac{N_2 - m}{3N} = \frac{(N_2-1)!}{3N^{N_2-1}}.
    \label{eq:1_to_N2}
\end{equation}
The logarithm of Eq.~\eqref{eq:1_to_N2} is approximated by Stirling's formula as follows:
\begin{equation}
    \ln\frac{(N_2-1)!}{3N^{N_2-1}} \approx N_2 \ln N_2 - N_2 - N_2 \ln N = -N_2 \left( 1-\ln \phi_2 \right) = - \left( 1-\ln \rho_2 \right) \phi_2 N.
    \label{eq:stirling}
\end{equation}
From $|N_2 \rangle$, the system can jump directly to the paradise when $a$ is chosen as the donor and one of $b_i$'s is chosen as the recipient ($i=1,\ldots,N_2$) with probability
\begin{equation}
    \frac{1}{N} \frac{N_2}{N} = \frac{\phi_2}{N}.
    \label{eq:rho2/N}
\end{equation}
Combining Eqs.~\eqref{eq:rho1_rho2}, \eqref{eq:stirling}, and \eqref{eq:rho2/N}, we conclude that the probability of transition from the initial segregation of $N_1$ versus $N_2$ to the paradise is bounded from below by
\begin{equation}
    T(\text{paradise} | N_1) \gtrsim
    C \frac{\phi_1 \phi_2^2}{N} e^{-(1-\ln \phi_2)\phi_2 N},
    \label{eq:Tparadise}
\end{equation}
where $C$ is a proportionality constant independent of $N$.
When multiplied by $q_{1'}^{-1} \sim (N-1)!$, Eq.~\eqref{eq:Tparadise} diverges as $N \to \infty$. Consequently, $P_{2\to 0} /q_{1'}$ should also diverge [see Eq.~\eqref{eq:P_2to0}].
According to Eq.~\eqref{eq:principal}, this divergence implies that the paradise will be the most probable outcome even if $N \to \infty$.

\setcounter{figure}{0}
\renewcommand{\thefigure}{E\arabic{figure}}

\section{L6 in sparse structures}
\label{app:husimi}

In the case of L6, the uniform stationary probability among balanced configurations is contrasted with a result from the Hamiltonian model defined by
\begin{equation}
H = -\sum_{\left< ijk \right>} \sigma_{ij} \sigma_{jk} \sigma_{ki}
\label{eq:ctd}
\end{equation}
because the paradise is locally stable in the Hamiltonian model~\cite{malarz2022mean}. Let us make sense of this difference: To define the Hamiltonian, every pair of vertices must be in a reciprocal relation, i.e., either mutually good or mutually bad, but it is approximately true in our setting as well because the convergence of $\sigma_{ij} \sigma_{ji} \to O(1)$ is a relatively fast process~\cite{oishi2021balanced}.
We have to look at the zero-temperature Hamiltonian dynamics, derived from Eq.~\eqref{eq:ctd} as
\begin{equation}
\sigma_{ij}' = \operatorname{sgn} \left( \sum_k \sigma_{ik} \sigma_{kj} \right),
\label{eq:tension}
\end{equation}
where the summation runs over the common neighbors of $i$ and $j$ except themselves.
Referring to those common neighbors is an important ingredient to ensure the local stability of the paradise because the peer pressure can correct an assessment error. In a sense, the Hamiltonian setting introduces \emph{surface tension} in such a way that the Ising model is contrasted with the voter model~\cite{dornic2001critical}.
The difference of Eq.~\eqref{eq:tension} from the L6 rule is evident in a large complete graph, but it vanishes if the underlying structure allows every pair of vertices to have only one common neighbor. The same applies even to the case of two common neighbors, if we interpret $\operatorname{sgn}(0)$ as $\pm 1$ with equal probabilities.

\begin{figure}
    \includegraphics[width=0.85\columnwidth]{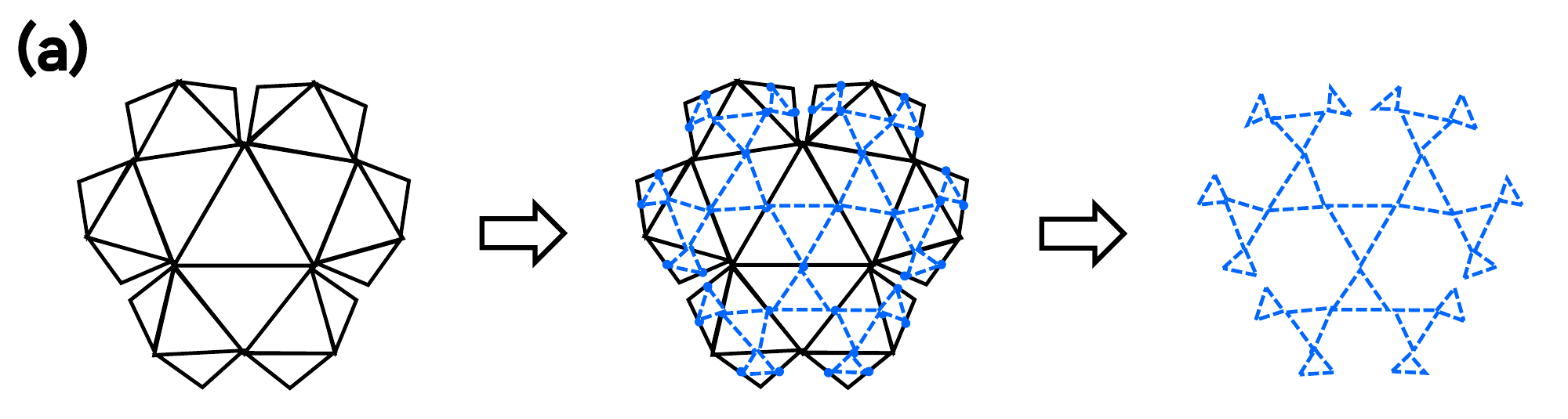}\\
    \includegraphics[width=0.85\columnwidth]{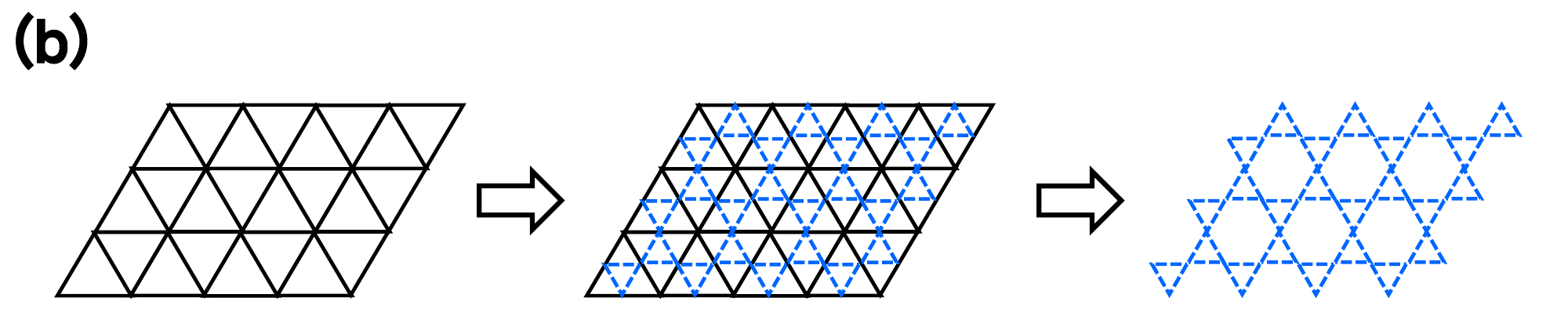}
    \caption{Line graphs obtained by transforming edges to vertices. (a) The line graph of the tree of triangles (solid) is the Husimi tree (dashed). (b) The line graph of the triangular lattice (solid) is the kagome lattice (dashed).}
    \label{fig:sparse}
\end{figure}

From the lack of directional preference in the transition among balanced configurations, we believe that the system governed by L6 with assessment error would be in a \emph{disordered} phase until reaching a balanced configuration. One possible mechanism to hinder a balanced configuration could be quenched random removal of edges, which corresponds to the case of $p<1$ in our model, where $p$ is the connection probability of a pair of vertices. When $p>O\left(N^{-1/2}\right)$ and $1-p > O\left(N^{-1}\right)$, the time to reach an absorbing state increases exponentially as $N$ grows~\cite{oishi2021balanced}, and this may well be a characteristic of a disordered phase.

To support this idea, let us consider a sparse structure as depicted on the left in Fig.~\ref{fig:sparse}(a). Each edge is shared by two triangles, so we expect that it can be studied by using the Hamiltonian approach as an approximation. Considering that our $\sigma_{ij}$'s are defined on edges, whereas spin variables are usually defined on vertices, we take the line graph of the original structure, which turns out to be the Husimi tree. The three-spin interaction Hamiltonian model on the Husimi tree is exactly solved~\cite{monroe1992phase,ananikian1997multifractal}, and the solution tells us that it is in a disordered phase. We have summarized the solution below for the sake of completeness.

Let us consider a system of Ising spins with the following Hamiltonian, defined on the Husimi tree:
\begin{equation}
    H=-J_3 \sum_{\langle i,j,k \rangle}\sigma_i \sigma_j \sigma_k -J_2\sum _{\langle i,j \rangle} \sigma_i \sigma_j - h\sum_i \sigma_i,
    \label{eq:Husimi_Hamiltonian}
\end{equation}
where $\langle i,j,k \rangle$ denotes a triangle of $i$,$j$, and $k$, and $\langle i.j \rangle$ means the nearest neighbors.
Let us denote the spin at the center by $\sigma_0$, and define the following function:
\begin{equation}
\begin{split}
    g_n(\sigma_0) &= \sum_{\{\sigma_1\}} \exp\Bigg[\tilde\beta\Bigg(J_3\sum_\Delta \sigma_0\sigma_1^{(1)} \sigma_1^{(2)} + J_2 \left(\sigma_0\sigma_1+ \sigma_0\sigma_2 +\sigma_1\sigma_2 \right) + h \sum _{j=1,2}\sigma_1^{(j)} \Bigg) \Bigg]\\
     &\times [g_{n-1}(\sigma_1^{(1)})]^{\gamma-1}[g_{n-1}(\sigma_1^{(2)})]^{\gamma-1},
    \end{split}
    \label{eq:g_n_Husimi}
\end{equation} $\\$
where $\tilde\beta \equiv 1/(k_B T)$, and $\gamma$ is the number of triangles at each vertex, which is two as depicted in Fig.~\ref{fig:sparse}(a). By using $g_n(\sigma_0)$, we can write the partition function as follows:
\begin{equation}
    Z = \sum_{\sigma_0} \exp(\tilde\beta h\sigma_0) [g_n(\sigma_0)]^{\gamma}
    \label{eq:Partition_function}
\end{equation}
Now, by defining $a \equiv e^{2\tilde\beta h}$, $b\equiv e^{2\tilde\beta J_2}$, and $c\equiv e^{2\tilde\beta J_3}$, we obtain the following map in terms of $z_n \equiv g_n(1)/g_n(-1)$~\cite{monroe1992phase}:
\begin{equation}
    z_n=y(z_{n-1}), \
    y(z) = \frac{a^2b^2cz^{2(\gamma-1)}+2az^{\gamma-1}+c}{a^2 z^{2(\gamma-1)} + 2acz^{\gamma-1}+b^2}
    \label{map_Husimi}
\end{equation}
The pure three-spin interaction Hamiltonian corresponds to $a=b=1$, in which case we have
\begin{equation}
    y(z) = \frac{cz^2+2z+c}{z^2 + 2cz+1}
    \label{map_Husimi_c}
\end{equation}
by setting $\gamma=2$. By solving $z=y(z)$, we obtain three fixed points, i.e., $z=-1$, $z=1$, and $z=-c$. However, $z=1$ is the only solution because $z$ cannot be negative. In addition, by drawing the map, we can see that $z=1$ is a stable fixed point.
The local magnetization is expressed as~\cite{monroe1992phase}
\begin{equation}
    \langle \sigma_0 \rangle =\frac{az_n^\gamma -1}{az_n^\gamma +1}
    \label{magnetization_Husimi},
\end{equation}
which is zero at $a=z=1$. However, the zero magnetization does not necessarily mean a disordered phase. To check the possibility of a phase transition, we will calculate the free energy. If it does not have a singularity at any finite temperature $T$, the system will always be in a disordered phase as in the high-temperature region. Let us rewrite the map by including both $a$ and $c$ as free parameters, while fixing $b=1$ and $\gamma=2$:
\begin{equation}
    y(z) = \frac{a^2cz^2+2az+c}{a^2 z^2 + 2acz+1}
    \label{eq:map_a_c}
\end{equation}
By rearranging the terms of $z=y(z)$, we obtain the following quadratic equation of $a$:
\begin{equation}
    (z^3-z^2c)a^2 + 2(z^2c-z)a +z-c=0,
    \label{eq:a_second_order_equation}
\end{equation}
which is solved by
\begin{equation}
    a = \frac{1-cz\pm \sqrt{1-c^2-z^2 +c^2z^2}}{z(z-c)}.
    \label{eq:a_solution}
\end{equation}
Noting that $a\ge 1$, $c\ge 1$, and $z>0$ by definition, the correct solution is given as follows:
\begin{equation}
    a(c,z) = \frac{1-cz - \sqrt{1-c^2-z^2 +c^2z^2}}{z(z-c)},
    \label{eq:z_range_a_solution}
\end{equation}
where $1\le z<c$. This is a monotonically increasing function of $z$, and this property will be used later. The free-energy density can be written as follows~\cite{baxter2007exactly}:
\begin{equation}
    \tilde\beta f = - \frac{2}{3}\tilde{J_3}-\tilde{h} - \int^{\tilde{h}}_{\infty} [M(\tilde{h}')-1]\ d\tilde{h}',
\end{equation}
where $\tilde{J_3}\equiv \tilde\beta J_3$ and $\tilde{h}\equiv \tilde\beta h$. The magnetic order parameter is denoted by $M = -\partial f/ \partial h$, which we may identify with $\langle \sigma_0 \rangle$ because of the translational symmetry of the lattice. The first two terms on the right-hand side is a constant of integration, which corresponds to the value in an ordered phase with $M=1$. We obtain $\tilde{h}$ from Eq.~\eqref{eq:z_range_a_solution} and differentiate it with respect to $z$ for $\partial \tilde{h}/\partial z$. We thus calculate the free-energy density as
\begin{equation}
    \tilde\beta f = - \frac{2}{3}\tilde{J_3}-\tilde{h} - \int^z_c [M(z')-1]\left[\frac{\partial \tilde{h}}{\partial z}\right]_{z=z'}\ dz',
\end{equation}
which indeed has no singularity at finite temperature. The conclusion is that the system will always be in a disordered phase as in the high-temperature region.

Another sparse structure that can be considered is the triangular lattice: In a Monte Carlo study, the Hamiltonian model on the triangular lattice has been reported to be disordered~\cite{malarz2020expulsion}. Its line graph is the kagome lattice [Fig.~\ref{fig:sparse}(b)], on which the exact solution of the three-spin Hamiltonian model again confirms that it is disordered~\cite{wu1989exact,barry2019finite}. All these observations indicate that a sparse structure would result in a disordered phase.

\setcounter{figure}{0}
\renewcommand{\thefigure}{F\arabic{figure}}

\section{Evolutionary aspects}

\setcounter{table}{0}
\renewcommand{\thetable}{F\arabic{table}}

\begin{table}
\caption{Leading eight. We denote cooperation and defection as $C$ and $D$,
respectively, and a player is assessed as either good ($G$) or bad ($B$).
To see what the assessment rule $\alpha_{uXv}$ means, let us assume that a donor and a recipient is observed by an observer, who assesses the donor and the recipient as $u$ and $v$, respectively. When the donor chooses $X\in \{C,D\}$ toward the recipient, the observer newly assesses the donor as $\alpha_{uXv} \in \{G,B\}$.
Likewise, the behavioural rule $\beta_{uv}$ tells the donor what to do between $C$ and $D$ if the donor has self-assessment $u$ and assesses the recipient as $v$.}
\begin{tabular}{c|cccccccc|cccc}\hline\hline
& $\alpha_{GCG}$ & $\alpha_{GDG}$ & $\alpha_{GCB}$ & $\alpha_{GDB}$ &
 $\alpha_{BCG}$ & $\alpha_{BDG}$ & $\alpha_{BCB}$ & $\alpha_{BDB}$ &
 $\beta_{GG}$ & $\beta_{GB}$ & $\beta_{BG}$ & $\beta_{BB}$\\\hline
L1 & G & B & G & G & G & B & G & B & $C$ & $D$ & $C$ & $C$\\
L2 & G & B & B & G & G & B & G & B & $C$ & $D$ & $C$ &
$C$\\
L3 & G & B & G & G & G & B & G & G & $C$ & $D$ & $C$ & $D$\\
L4 & G & B & G & G & G & B & B & G & $C$ & $D$ & $C$ & $D$\\
L5 & G & B & B & G & G & B & G & G & $C$ & $D$ & $C$ & $D$\\
L6 & G & B & B & G & G & B & B & G & $C$ & $D$ & $C$ & $D$\\
L7 & G & B & G & G & G & B & B & B & $C$ & $D$ & $C$ & $D$\\
L8 & G & B & B & G & G & B & B & B & $C$ & $D$ & $C$ & $D$\\\hline\hline
\end{tabular}
\label{tab:eight}
\end{table}

By norm, here we always mean
a combination of an assessment rule and an action rule. This was indeed
the case in the first report on the leading eight~\cite{ohtsuki2004should},
according to which both of L4 and L6 are defined with a
common action rule. What the authors of Ref.~\onlinecite{ohtsuki2004should}
showed is that only eight norms, i.e.,
eight combinations of assessment and action rules, successfully
resist the invasion of mutants with alternative action rules. As long as
we restrict ourselves to the
class of norms whose assessment is based solely on the last observation,
therefore, even if we decouple assessment and action, all the other norms except
the eight combinations will
be removed if we apply their evolutionary screening criterion.
For this reason, we have followed the original definition of the leading eight,
specified by \emph{both}
the assessment and action rules \skb{(Table~\ref{tab:eight})}.

We have proposed L4 as a remedy for segregation induced by L6.
When it comes to evolution of cooperation, one way to justify this proposal
could be to separate time scales between the mutation of action rules
and that of assessment rules.
We argue the reason as follows:
In Ref.~\onlinecite{perret2021evolution}, the authors have analyzed
the invasion of mutants with alternative assessment rules and concluded that
Always Defect (AllD) is the only evolutionarily stable strategy when error is
not negligible. Although their analysis has left some room for further
investigation of more general actions rules, we regard their overall conclusion
as plausible. It is also qualitatively
consistent with the failure of L4 as reported in
Ref.~\onlinecite{hilbe2018indirect},
where L4 showed poor performance when
competing with Always Cooperate (AllC) and AllD.
To obtain something other than full defection, therefore, we may
consider restricting the set of available norms to the leading eight, e.g., by
applying the screening procedure in Ref.~\onlinecite{ohtsuki2004should},
under the
assumption that the mutation of action rules occurs constantly so that only the
leading eight can survive with significant frequencies.

\begin{figure}
{\centering
\includegraphics[width=0.45\textwidth]{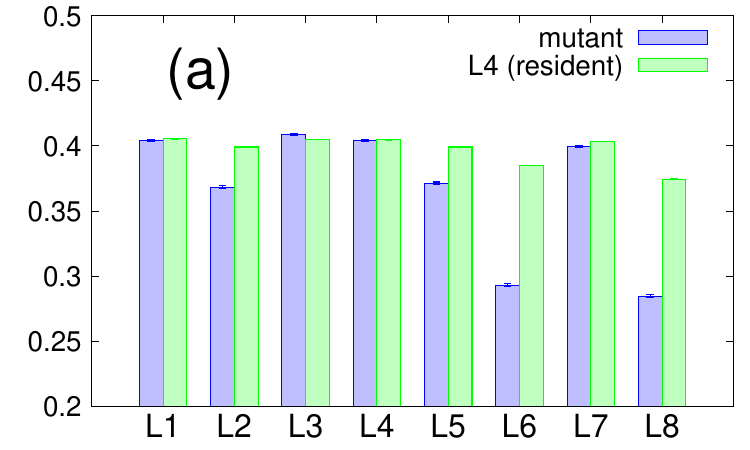}
\includegraphics[width=0.45\textwidth]{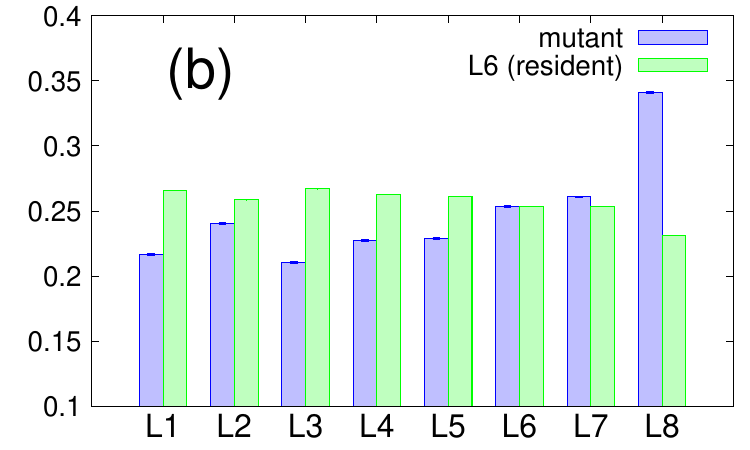}
\par}
\caption{Average individual payoffs when the resident norm is being invaded
by each of the leading eight. The resident norm is L4 in (a) and L6 in (b).
The population size is $N=50$,
among which the fraction of mutants is $10\%$. At each time step, two randomly
chosen players interact by playing the donation game, in which the benefit and
the cost of cooperation are set to be $b=1$ and $c=1/2$, respectively.
Assessment error occurs with probability $5\%$, and the same is also true for
execution error. We have discarded the first $T=1.5 \times 10^6$ time steps to
get stationary results. This graph shows how much each individual obtains
between $T$ and $2T$ steps on average, and we have repeated such simulation $10$
times to estimate the error bars.
}
\label{fig:invasion}
\end{figure}

\begin{figure}
{\centering
\includegraphics[width=0.45\textwidth]{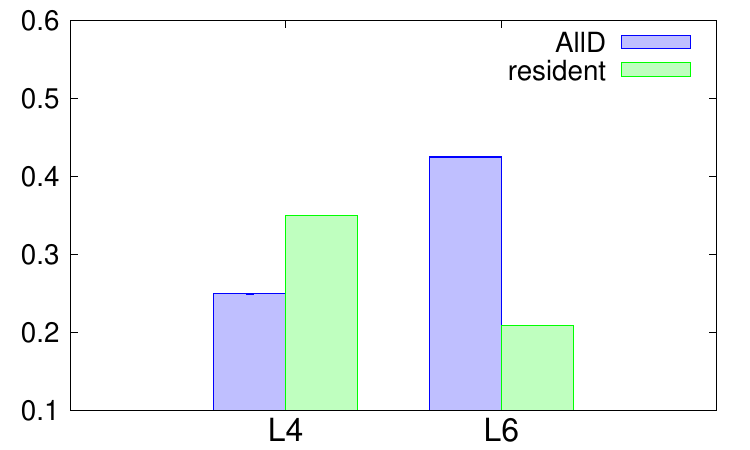}
\par}
\caption{Average individual payoffs when the resident norm is either L4 or L6, whereas the mutant norm is AllD, defined by $\alpha_{uXv}=B$ and $\beta_{uv}=D$ regardless of $u$, $X$, and $v$. The other parameters for the simulations, including the error probabilities, are the same as in Fig.~\ref{fig:invasion}.
}
\label{fig:alld}
\end{figure}

Now, let us suppose that the invasion analysis is carried out among the
leading eight with an initial condition that the resident population uses L4.
Then, our numerical calculation shows that the resident population is clearly
better off than L2, L5, L6, or L8 mutants, and almost selectively neutral to L1,
L3, or L7 mutants [Fig.~\ref{fig:invasion}(a)]. This implies that L4 can remain
the governing norm for a substantial amount of time, in which case
our present work shows that the system will almost surely
reach the paradise where cooperation abounds.
Interestingly, when the mutation of action rules is a fast process, leaving the
assessment rule approximately constant, we approach the situation described in
Ref.~\onlinecite{santos2021complexity}. However, the difference is that their work uses the
public-reputation model in which an erroneous assessment is also shared by all.
In the private-reputation model, by contrast, L4 is known to be more cooperative than L6
(see, e.g., Ref.~\onlinecite{fujimoto2024leader}).
For the sake of completeness, we have additionally checked the stability of L4 and L6 against AllD, which judges everyone as bad and prescribes defection all the time unless error occurs. Our numerical calculation shows that L4 is better off than AllD, whereas L6 is invaded by AllD (Fig.~\ref{fig:alld}), in accordance with the findings of Ref.~\onlinecite{fujimoto2024leader}.

\begin{figure}
{\centering
\includegraphics[width=0.45\textwidth]{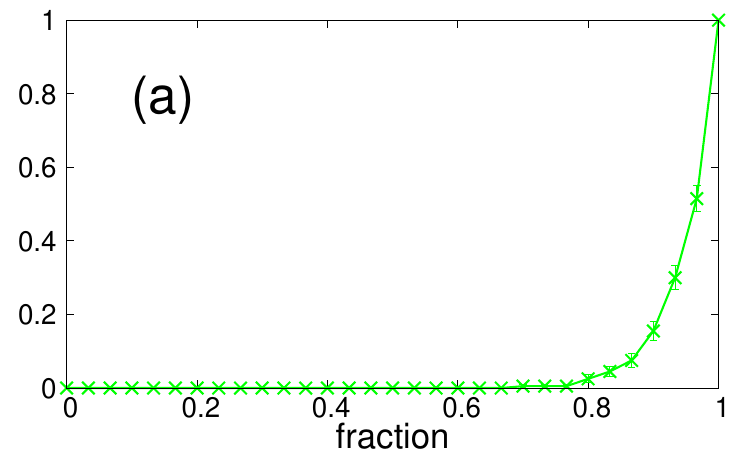}
\includegraphics[width=0.45\textwidth]{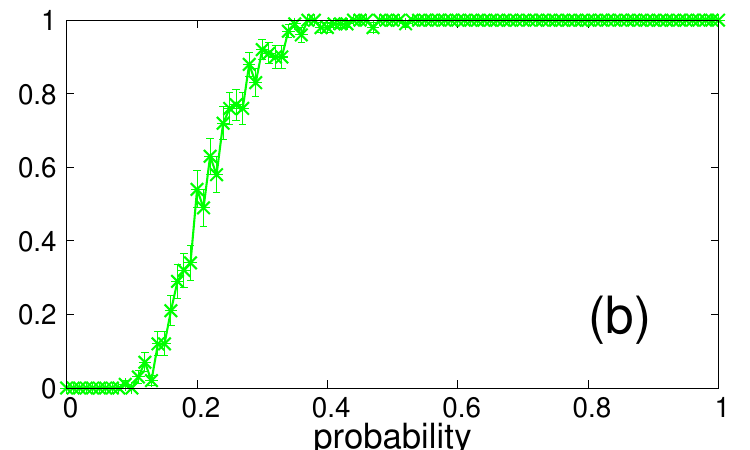}
\par}
\caption{Probability of reaching the paradise when L4 and L6 coexist. (a) The
horizontal axis means the fraction of individuals playing L4, when the rest of
the population use L6. (b) Each individual uses a mixed strategy of L4 and
L6, and the horizontal axis means the probability of using L4. The population
size is $N=30$, and we have averaged $10^2$ independent samples to get each data
point.
}
\label{fig:mix}
\end{figure}

We may also ask ourselves a couple of other questions. One is whether a small number of L4 players can invade a resident population of L6. From
Fig.~\ref{fig:mix}(a), one could guess that they can, but it is not the
case: When L6 is the majority, it
is better off than L4 mutants [Fig.~\ref{fig:mix}(b)].
In other words, strong frequency-dependent selection is
working between L4 and L6, and the result is that whoever the majority is rules.
We thus conclude that L4 would need certain critical mass to achieve fixation
directly in an L6 population.
Another related question is then how much it helps when a small number
of resident players convert from L6 to L4.
Figure~\ref{fig:mix}(a) shows that more than
$80\%$ of the population have to convert for making a visible
change in the probability of reaching the paradise. The situation becomes more
optimistic if everyone partially adopts L4 and uses it with certain probability.
In this case, as soon as the probability of using L4 exceeds about $10\%$,
we observe a clear increase in the probability of reaching the paradise
[Fig~\ref{fig:mix}(b)].

\setcounter{figure}{0}
\renewcommand{\thefigure}{G\arabic{figure}}

\section{Effects of execution error}

In the main text, we have considered assessment error only. One of the reasons is that
execution error has only limited effects on the dynamics without altering
the overall picture.
To see this, let us begin with a balanced configuration of a population governed by the
following L6 rule:
\begin{equation}
\sigma_{od}' = \sigma_{or} \sigma_{dr}
\label{eq:L6'}
\end{equation}
We assume that a player, say, Alice, defects against her
friend by committing an execution error. All her friends (as well as Alice
herself) judge it as bad, whereas her enemies do the opposite. Now, as long as
everyone follows L6 exactly, this configuration will not change, unless Alice is
chosen as a donor once again. This statement is obviously true when Alice is a
mere observer because how people assess her is irrelevant to this situation.
When she is a recipient, both $\sigma_{or}$ and $\sigma_{dr}$ take the opposite
signs compared to the original balanced configuration, leaving the left-hand
side of Eq.~(\ref{eq:L6'}) unchanged. If Alice becomes a donor again, by choosing
the correct action, she regains the original assessments from both her friends
and foes. The same kind of reasoning can be made when it comes to erroneous
defection as well. In short, the system has no other way than to return back to the original
balanced configuration after an execution error.

\begin{figure}
{\centering
\includegraphics[width=0.45\textwidth]{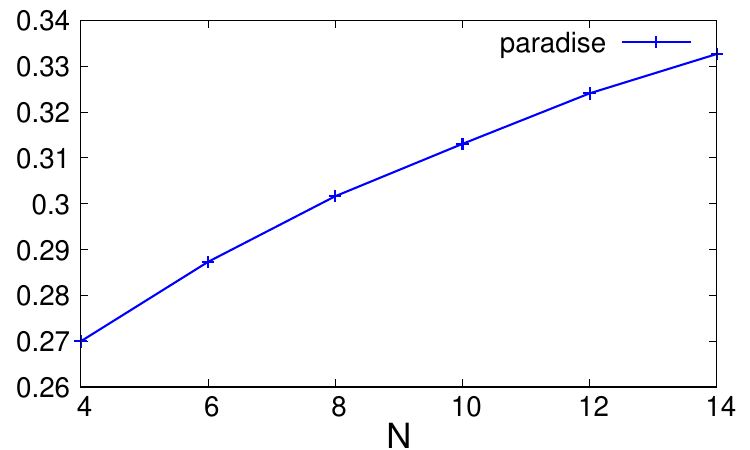}
\par}
\caption{Effects of execution error in L4 when the initial balanced
configuration consists of two equal-sized clusters. Our Monte Carlo simulations
shows that the system either goes to the paradise or comes back to the original
balanced configuration. This plot shows the probability to go to the paradise,
which increases as the system size $N$ grows. Each data point is an estimate from
$10^5$ independent samples.}
\label{fig:Exe_L4_paradise}
\end{figure}

In the case of L4, the difference is that Alice would not lose her good
reputation among her friends even after cooperating toward one of her
foes by mistake. The
system either comes back to the original balanced configuration as we just saw
in the case of L6, or moves directly to the paradise. Once it arrives at the
paradise, it is impossible to escape from it by execution error.
Figure~\ref{fig:Exe_L4_paradise} shows our Monte Carlo estimates of the
conditional probability of reaching the paradise after an execution error when
the initial balanced configuration consists of two equal-sized clusters. As the
system size $N$ increases, it becomes more probable to end up at paradise.

%

\end{document}